\documentclass[twoside]{article}
\usepackage{cite}
\usepackage{epsf}
\usepackage{epic}
\usepackage{latexsym}
\usepackage{amsfonts}

\columnwidth\textwidth

\begin{document}

\thispagestyle{empty}
\parskip=12pt
\raggedbottom

\begin{flushright}
BUTP--97/16
\end{flushright}

\vspace*{2cm}

\begin{center}
{\LARGE Properties of Renormalization Group Transformations
\footnote{Work supported in part by Schweizerischer Nationalfonds}}

\vspace*{1cm}

Peter Kunszt \\
Institut f\"ur theoretische Physik \\
Universit\"at Bern \\
Sidlerstrasse 5, CH--3012 Bern, Switzerland

\vspace*{0.5cm}

{June 1997} 

\vspace*{3cm}

\nopagebreak[4]

\begin{abstract}
We describe some properties of Renormalization Group
transformations. Especially we show why some of the RG
transformations have redundant eigenoperators with eigenvalues that cannot be
determined by simple dimensional analysis and give the corresponding formulae.
\end{abstract}

\end{center}

\newpage
\setcounter{page}{1}


\section{Introduction}

There is an ongoing effort to find improved lattice actions for asymptotically
free theories like QCD that are nearly  free of lattice artifacts by
the means of renormalization group theory \cite{ym,fermqcd}. The conceptual
basics have been developed by Hasenfratz and Niedermayer for the $O(3)$
non-linear $\sigma$-model \cite{sigmamod}. A good description of the basic
ideas of improved lattice actions can be found in \cite{ferenc}.  While using
the renormalization group transformation (RGT) 
to get improved actions, it is important to understand
the basic convergence properties of the transformation in the vicinity of the
fixed point. In the standard way of thinking, the eigenvalues of the RG
transformation can be determined by dimensional analysis. However, general RG
transformations result in additional non-canonical eigenvalues.
In this paper we give an insight how this mechanism works.
Here we consider free fermionic field theories although the statements are
more general. 
 
The standard terminology is the following: For fermions, when we
transform a fine lattice to a coarse one, we have 
\begin{equation}
e^{ -S^\prime[\bar{\Psi}^\prime,\Psi^\prime]} = \int{\cal D}\bar{\Psi}\,{\cal
  D}\Psi\, e^{-S[\bar{\Psi},\Psi] -
  T[\bar{\Psi},\Psi;\bar{\Psi}^\prime,\Psi^\prime]}
\label{blocktrafo} 
\end{equation} 
where $\bar{\Psi},\Psi$ and $\bar{\Psi}^\prime,\Psi^\prime$ are the fermion fields on the fine
and coarse lattice and $T$ defines the renormalization group or block
transformation that does the averaging. In principle, there are no
restrictions to $T$. But we want to choose an averaging that matches our
intuitive picture of the renormalization group transformation. That means that
we do not want $T$ to average over too much or too few sites on the fine
lattice, just enough to account for the space between the coarse lattice
points.  It turns out that the interaction range of the FP action depends
significantly on the actual choice of $T$.

Apart from these qualitative restrictions, one of the 
constraints that applies to $T$ is that the generating functional $Z$
(and thus the free energy and thermodynamics) should not change:
\begin{equation}
Z \;=\; \int{\cal D}\bar{\Psi}\,{\cal D}\Psi\,
\exp\left\{-S[\bar{\Psi},\Psi]\right\} \quad =\; \int{\cal D}\bar{\Psi}^\prime\,{\cal D}\Psi^\prime\,
\exp\left\{-S^\prime[\bar{\Psi}^\prime,\Psi^\prime]\right\}
\end{equation}
The long-distance properties are thus untouched. The correlation length
measured in lattice spacings (the dimensionless counterpart of the
physical correlation length; $\xi = \xi^{\mbox{\tiny phys}}/a$) is
changed to $\xi/n$ since the lattice spacing is changed $a^\prime=na$ and
thus $\xi^\prime = \xi/n$. In this work we only use transformations with $n=2$.

\section{The RGT, its Operators and Fixed Points}
\subsection{Eigenoperators and Eigenvalues of the RGT}\label{eigencucc}

The action $S^\prime[\bar{\Psi}^\prime,\Psi^\prime]$ is most generally a sum of all kinds of
interaction terms even if $S[\bar{\Psi},\Psi]$ had a simple form. If we set
the initial $S$ to the general form
\begin{equation}
S[\bar{\Psi},\Psi] = -\sum_i c_i X_i[\bar{\Psi},\Psi] \label{genericS}
\end{equation}
where $c_i$ denotes the couplings of the different interactions
$X_i[\bar{\Psi},\Psi]$ ($i$ is just a labeling index). The interactions may
be of any form conformal to lattice symmetries. These can be higher
derivatives and higher powers of the fields $\bar{\Psi},\Psi$. The couplings
$c_i$ form a coupling space. An RG transformation is moving the action
in this space, and for very special sets of couplings $c_i^\ast$ the RG
reproduces the same couplings -- we have a fixed point FP of the RGT. 
In such FPs we have collective behavior over
infinite many lattice sites. In the vicinity of the FP-set of couplings
$\{c_i^\ast\}$ the function $c_i^{(n+1)}(c_j^{(n)})$ is given by
\begin{eqnarray}
c_i^{(n+1)}(c_j^{(n)})&\!\!\!\! =\!\!\!\! &c_i^{(n+1)}(c_j^{\ast})\!+\!
\frac{\partial c_i^{(n+1)}}{\partial c_j^{(n)}}\!\!\left(\!c_j^{(n)}\!\!-c_j^{\ast}
\!\right)\!+\!\cdot\!\!\cdot\nonumber\\
\label{expfp}
\end{eqnarray}
So because $c^\ast$ is a FP, $c_i^{(n+1)}(c_j^\ast)=c_i^\ast$ we have
\begin{equation}
\Delta c_i^{(n+1)} = R_{ij} \Delta c_j^{(n)}\label{linRGT}
\end{equation}
where the $\Delta$ denotes the difference of the point $c$ to
$c^\ast$.  The procedure that follows is the usual one for a
linearized symmetry: we try to identify its eigenvectors and
eigenvalues. Denote the eigenvalues with $e_\alpha$ and the
eigenvectors with $\sigma_\alpha$. The eigenvectors $\sigma_\alpha$
are some linear combinations of the operators
$X_i$. The neighborhood of the FP can be then expressed as
\begin{equation}
S[\bar{\Psi},\Psi] = S^\ast[\bar{\Psi},\Psi] + \sum_\alpha
d_\alpha\sigma_\alpha[\bar{\Psi},\Psi]\label{fixpointneighborhood}
\end{equation}
Performing one RG step we get
\begin{equation}
S^\prime[\bar{\Psi},\Psi] = S^\ast[\bar{\Psi},\Psi] + \sum_\alpha e_\alpha
d_\alpha\sigma_\alpha[\bar{\Psi},\Psi] \label{nearfp}
\end{equation}
If we continue to apply the blocking step to this action, we will get an
additional factor $e_\alpha$ for each step. Of course this action will
only converge towards $S^\ast$ if every eigenvalue $e_\alpha$ in the sum
is smaller than $1$. Eigenvectors with such eigenvalues are called
irrelevant. The most general action $S$ will have of course
eigenvector-components with (relevant) eigenvalues larger than $1$
and thus the repeated application of
the RG transformation will bring the system away from the FP. In order
to converge to the FP, the coefficients $d_\alpha$ have to be set to
zero for every relevant eigenvector $\sigma_\alpha$.

\subsection{Redundant Operators}\label{redundantoperators}

For the Gaussian Fixed Point, one can give the form of the eigenoperators and
eigenvalues explicitly for specific RG transformations. 
The eigenvalues are depending of the engineering dimensions of the
operators involved. They are all some exponent
of $2$: We have $e_\alpha=1,\frac{1}{2},\frac{1}{4},\frac{1}{8},\ldots$

If we have some general RG transformation, one would think that the
eigenvalues should be reproduced, and that specifically there is no
eigenoperator with an eigenvalue $1/2<e_\alpha<1$ because it is
impossible to write a regular operator with such an engineering
dimension.

What we are seeing in the calculations, however, is that we do measure
such eigenvalues. These eigenoperators with an eigenvalue between $1/2$
and $1$ are interpreted as redundant directions in the critical surface,
i.e.~connecting lines between possible FP actions. That's why these
eigenvalues have no physical significance. To be specific, we move on
such a direction from a FP1 of a RG transformation (RG1), to FP2 of some
other RG transformation (RG2) which is very similar to RG1. The
'difference' between one transformation and the other one is giving the
redundant operator and the eigenvalue may be in principle any number. If
we are in the FP1 of RG1 and now apply RG2, we are moving towards FP2
with such a redundant eigenvalue. If this eigenvalue is smaller than
$1$, we will eventually end up in FP2. If this eigenvalue is greater or
equal $1$, this means that we cannot move from FP1 to FP2 since FP1 does
not lie in the attracting region of FP2. In
particular, if the eigenvalue is exactly $1$, then the points on the
line between FP1 and FP2 will not move under RG1 or RG2 at all -- the FPs of
these transformations are not unique. This fact does not matter for
continuum physics, because every point on the connecting line leads to
an action which is describing the same theory.

Operators belonging to such directions are redundant, their presence
in the action does not influence the lattice artifacts. Those are only
influenced by the irrelevant operators.

\subsection{The Blocking Function $T$ and the FP}

Consider some lattice regularization of the maselss free fermionic theory: 
\begin{eqnarray}
S[\bar{\Psi},\Psi] &=& \sum_{q}\bar{\Psi}(q)\,\Delta(q)\,\Psi(-q)\label{Sdef1}\\
\Delta(q) &=& \rho_\mu(q)\gamma_\mu + \lambda(q)
\end{eqnarray}
Usually,  $T$ is written as
\begin{eqnarray}
T[\bar{\Psi},\Psi;\bar{\Psi}^\prime,\Psi^\prime]\; = \; \hspace*{10cm}\nonumber \\
\kappa\,\sum_{n_{\scriptscriptstyle \!\! B}}\left(\bar{\Psi}^\prime(n_{\scriptscriptstyle \!\! B})-b\,
\sum_{j}\, \omega^\dagger(j)\bar{\Psi}(2n_{\scriptscriptstyle \!\! B}+j)\right)\left(\Psi^\prime(n_{\scriptscriptstyle \!\! B})-b\,
\sum_{j}\, \omega(j)\Psi(2n_{\scriptscriptstyle \!\! B}+j)\right)\label{Tdef}
\end{eqnarray}
The function $\omega(j)$ is normalized and determines the exact form of the
block transformation ( $\sum_j\omega(j)=1$ ). 
The integral over $\bar{\Psi}$ and $\Psi$ in (\ref{blocktrafo}) is a Gaussian
integral and it is easily carried out. 
The action after one RG transformation has the form
\begin{equation}
S^\prime[\bar{\Psi}^\prime,\Psi^\prime]= \sum_{Q}\bar{\Psi}^\prime(Q)\,\Delta^\prime(Q)\,\Psi^\prime(-Q)\label{Sonestep}\\ 
\end{equation}
\begin{equation}
\Delta^{\prime -1}(Q) =
\frac{1}{\kappa}+\frac{b^2}{2^d}\sum_{\ell=0}^{1}
\Delta^{-1}\left(\frac{Q}{2}+\pi\ell\right)\left|\,
\omega\left(\frac{Q}{2}+\pi\ell\right)\right|^2
\label{momdelta}
\end{equation}
This is our new effective action for the coarser lattice. The initial
action can be in principle any fermionic action of the same quadratic
form. 
Here $Q$ is the coarse-lattice momentum and $d$ counts the dimensions
(usually $d=4$).  This expression is very convenient for iterative
purposes. 
The iteration can be carried out for the inverse of the action kernel
(i.e.~the propagator). Since $\Delta(q)$ has a simple matrix structure
(\ref{Sdef1}) the inversion $\Delta^{-1}(q)$
to get the propagator is easy and the recursion straightforward, giving the
same matrix structure $(\alpha_\mu\gamma_\mu + \beta{\bf 1})$.
Having found a recursion relation for the propagator $\Delta^{-1}$, 
we can invert it back to get the new action kernel.
Then we can iterate the action $S$ to obtain the fixed point action $S^\ast$.
After $k$ iterations, the propagator functions $\alpha^{(k)}_\mu(Q)$ and
$\beta^{(k)}(Q)$ are given by
\begin{eqnarray} \alpha_\mu^{(k)}(Q)\!\! &\!\! =\!\! &\!\!\frac{1}{2^k}\!\!
\sum_{\ell=0}^{2^k-1}\!\alpha_\mu\!\left(\!
  \frac{Q+2\pi\ell}{2^k}\right)\!\!\left|\Omega^{(k)}(Q+2\pi\ell)\right|^2\label{alphak}\\
\nonumber\\
\beta^{(k)}(Q)\!\! &\!\! =\!\!
&\!\!\frac{1}{\kappa}\,+\,\frac{1}{\kappa}\,\sum_{n=1}^{k-1}\, \frac{1}{2^n}
\sum_{\ell=0}^{2^n-1}\,\left|\Omega^{(n)}(Q+2\pi\ell)\right|^2 +\, \frac{1}{2^k}\sum_{\ell=0}^{2^k-1}\,\beta\left(
  \frac{Q+2\pi\ell}{2^k}\right)\,\left|\Omega^{(k)}(Q+2\pi\ell)\right|^2
\hspace*{2mm}\label{betak}
\end{eqnarray} 
where 
\begin{equation} \Omega^{(k)}(Q)=\prod_{i=1}^k\,\,
\omega\left(\frac{Q}{2^i}\right)\label{bigomega} 
\end{equation} 
The number $b$ is has been fixed to $b=2^{(d-1)/2}$.
For other values, $\alpha_\mu$ either diverges or goes to
zero.  For large $k$ in eqs.~(\ref{alphak},\ref{betak}) only the small
momentum behavior of the original propagator $\alpha_\mu(q)\gamma_\mu +
\beta(q)$ matters. Starting the iteration from the Wilson fermion action
$\rho_\mu(q) = \sin q_\mu$; $\lambda(q) = m + 2r\sum_\mu\sin^2 (q_\mu/2)$
(where $r$ is just a free parameter) we get
\begin{eqnarray}
\alpha_\mu^{(k)}(Q;m)\!\! &\!\! =\!\! &\!\!\!\!
\sum_{\ell=0}^{2^k-1}\!\!
\frac{Q_\mu+2\pi\ell_\mu}{(Q+2\pi\ell)^2\!+\!m^2}\left|\Omega(Q+2\pi\ell)\right|^2\label{rtalpha}\nonumber\\
\\
\beta^{(k)}(Q;m)\!\! &\!\! =\!\! &\!\! \frac{1}{\kappa}\;\hat\omega(Q)
\,+\,\sum_{\ell=0}^{2^k-1}
\, 
\frac{m}{(Q+2\pi\ell)^2+m^2}\;\left|\Omega(Q+2\pi\ell)\right|^2\label{rtbeta}
\end{eqnarray}
with $m=2^kM$. In this relation we see that $M$ is the one and only
relevant parameter of the RG transformation for the free fermion field:
it is the only quantity that is doubled with each step. The only value
of $M$ which can lead to a fixed point is therefore $M=0$. This fixes
the so-called critical surface for the parameters of the free fermion
action.
In the infinity limit we get finally the FP quantities
\begin{eqnarray}
\alpha_\mu^\ast(Q)& =&
\sum_{\ell\in\mathbb Z}\,
\frac{Q_\mu+2\pi\ell_\mu}{(Q+2\pi\ell)^2}\;\left|\Omega(Q+2\pi\ell)\right|^2\label{fpalpha}
\\
\beta^\ast(Q)&=&\frac{1}{\kappa}\,\hat\omega(Q)\label{fpbeta}
\end{eqnarray}
where
\begin{eqnarray}
\hat\omega(Q)&=&1+\sum_{n=1}^{\infty}\,
\frac{1}{2^n}
\sum_{\ell=0}^{2^n-1}\,\left|\Omega^{(n)}(Q+2\pi\ell)\right|^2\label{defomegahat}
\end{eqnarray}
For the sake of convergence, we have the (weak) conditions on the
blocking function $\omega$ that the infinite sum $\hat\omega$ and the
infinite product $\Omega$ must be finite functions of $Q$ and that the
summation over $\ell$ is also leading to a finite value. As an
additional constraint, we have the normalization condition
$\omega(Q=0)=1$. The summation over $\ell$ in (\ref{fpalpha}) is now
over all positive and negative numbers because due to periodicity
($f(-Q)=f(2\pi-Q)$) the summation $\ell = 0,\ldots,2^k-1$ in
(\ref{alphak}) may be rewritten to $\ell = -2^{k-1}+1,\ldots
2^{k-1}$. This form is more convenient in the $k\rightarrow\infty$ limit
because terms with $\ell = \pm|\ell|$ are equally important. A fixed
point exists if the infinite sums in (\ref{fpalpha},\ref{fpbeta})
converge.
The number $\kappa$ has not been fixed by convergence considerations
in the above section. So the given RG transformation may be
parametrized by $\kappa$. 
\section{Examples of Block Transformations}
\subsection{Decimation}
Decimation is the crudest method of blocking the lattice. It consists
of simply keeping only one site in a block and integrating out the rest:
\begin{equation}
e^{-S^\prime[\bar{\Psi}^\prime,\Psi^\prime]}=\int{\cal
  D}\bar{\Psi}\,{\cal D}\Psi\prod_{n_{\scriptscriptstyle \!\! B}}\delta(\bar{\Psi}^\prime(n_{\scriptscriptstyle \!\! B})-\bar{\Psi}(2n_{\scriptscriptstyle \!\! B}))
\cdot\delta(\Psi^\prime(n_{\scriptscriptstyle \!\! B})-\Psi(2n_{\scriptscriptstyle \!\! B}))\cdot
\exp\left\{-S[\bar{\Psi},\Psi]\right\}
\end{equation}
Writing this in terms of eqs.~(\ref{blocktrafo},\ref{Tdef}) this
corresponds to $\omega(j)=\delta_{j,0}$ and $\kappa=\infty$
We have to check if this crude transformation meets our conditions for
the FP to exist. In one dimension we find that there is a fixed point
of the RG transformation and we can give the explicit form of the
FP. Starting from equations (\ref{alphak},\ref{betak}) with
$\Omega(Q)=1$ and $m=0$ we get $\alpha^D(Q)=1/2\cot(Q/2)$ and $\beta^D(Q)=r/2$
So that the action is given by the functions
\begin{eqnarray}
\rho^D(Q)&\;=\;&\frac{\sin(Q)}{1+(r^2-1)\sin^2(Q/2)}\label{decrho}\\
\lambda^D(Q)&\;=\;&\frac{2r\sin^2(Q/2)}{1+(r^2-1)\sin^2(Q/2)}\label{declambda}
\end{eqnarray}
This action is the result of a sequence of infinitely
many RG steps, so it describes the continuum as any other FP
action. There is one peculiarity that needs attention. The FP has an
explicit dependence on the Wilson parameter $r$. So there are different
FPs for different starting actions. As a consequence, the parameter $r$
must parametrize FPs lying on a redundant direction with eigenvalue $1$,
as discussed in section \ref{redundantoperators}. This RG transformation
gives a line of FPs, so it cannot move from one FP to another of this
line -- each FP will block into itself. The explicit dependence on a
free parameter should thus not bother us, it is just another parameter
that may be tuned for other properties of the action.

We have stated above that the FP is only depending on the exact form of
the RG transformation as a consequence of the basic hypothesis that the
long ranged interactions are independent of the exact form of the
short-ranged fluctuations. That the FP depends on another free parameter
is only stating that we may end up in different FPs for the same
transformation. This case may be regarded as a limiting case to where no
FP exists (the RG transformation diverges). 

For $r=1$ the decimation-FP action is just the Wilson-fermion
action. Unfortunately in one dimensions we have no spectrum $E(q)$
so we cannot check for the only physical quantity in the free
system.

Decimation is not a good RGT because it represents the fine lattice
action {\em too} well -- the correlation functions on the coarse and fine
lattice are identical.
So even after an arbitrary high number of RG steps the action on the
final coarse lattice will depend on the microscopic details of the
starting action, it does not `forget' anything. This is of course not
what we want. An RGT should average out the microscopic details in order
to concentrate on the long-range behaviour. Decimation simply copies
everything to a larger lattice.

In higher dimensions, the FP formula (\ref{fpalpha}, \ref{fpbeta})
diverges -- there is too much to keep. The
FP with an explicit dependence on a parameter is not a
speciality for one dimension, we could also give a block transformation which
has a fixed point with an explicit $r$-dependence in two or more dimensions.

\subsection{Kadanoff-blocking}

As opposed to decimation, the blocking procedure presented by Kadanoff
takes all sites of the fine lattice into account.  Their weight is
equal within the block.  The procedure is simple: Always block $2^d$ points
together by building their average:
\begin{equation}
\omega(j) = 2^{-d}\sum_{n=0}^{2^d-1}\delta_{j, u^{(n)}}
\end{equation}
where $u^{(n)}$ are vectors pointing to the corners of the unit
cube; their components $u_i^{(n)}$ are either 1 or 0.

For this simple case, it is possible to calculate $\Omega^{(k)}(Q)$ and
$\hat\omega(q)$ explicitly \cite{wilsonbell}. From (\ref{bigomega}) we have
\begin{equation}
\Omega^{(k)}(Q)=\prod_\nu
\left(\frac{\sin(Q_\nu/2)}{2^k\sin(Q_\nu/2^{k+1})}\right)^{\!\!\! 2}
\label{wilsOmega}
\end{equation}
Using this in the definition of $\hat\omega(Q)$ in (\ref{defomegahat})
we see that the sum over $\ell$ gives just 1 for all n as a special
property of this blocking. This leaves us with a geometrical series
for $1/2$ giving for $k\rightarrow\infty$
\begin{equation}
\hat\omega(Q)=2
\end{equation}
The expression for $\Omega(Q)$ in the $k\rightarrow\infty$ limit is
very simple:
\begin{eqnarray}
\alpha_\mu^\ast(Q)&\!\! =\!\! &\!\!
\sum_{\ell\in\mathbb Z}\,
\frac{Q_\mu+2\pi\ell_\mu}{(Q+2\pi\ell)^2}\;\prod_\nu
\left(\frac{\sin(Q_\nu/2)}{\frac{Q_\nu}{2}+\pi\ell_\nu}
\right)^{\!\!\! 2}
\label{wilsalpha}\\
\nonumber\\
\beta^\ast(Q)&\!\!=\!\!&\!\!\frac{2}{\kappa}\label{wilsbeta}
\end{eqnarray}

This result was already presented in \cite{wiese}. In this case
the FP action may be given even analytically.

\subsection{Overlapping block transformation}

A more general blocking procedure is to average not just every
distinct unit cube as in the Kadanoff-blocking but to build the weighted
average of the fine points surrounding the blocked lattice site. This
way we may also consider blockings somewhere between decimation and the
Kadanoff-style flat blocking.
Explicitly, we can write for $d$ dimensions

\begin{equation}
\omega(j)=\left\{ \begin{array}{ll}
                c_o & j=0\\
                c_1 & j=\pm\hat{\mu}\\
                c_2 & j=\pm\hat{\mu}_1\pm\hat{\mu}_2 \hspace*{10mm} \mu_1\neq\mu_2\\
                c_3 & j=\pm\hat{\mu}_1\pm\hat{\mu}_2\pm\hat{\mu}_3 \;\; \mu_i\neq\mu_j\\
                \vdots\\
                c_d & j=\pm\hat{\mu}_1\pm\cdots\pm\hat{\mu}_d=\\
&\qquad (\pm 1,\ldots,\pm 1) \;\;\;\;  \mu_i\neq\mu_j\\
                0   & \mbox{for all other $j$}
                \end{array}
\right.\label{ovblock}
\end{equation}

\begin{figure}[ht]
\begin{center}
\leavevmode
\epsfxsize=60mm
\epsfbox{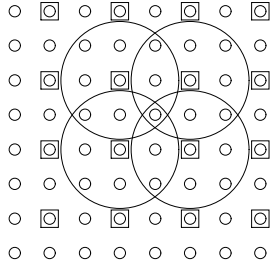}
\end{center}
\caption{\em Illustration of the overlapping: Unlike in the Kadanoff
blocking, one fine lattice point (circles) contributes to more than
one blocked lattice site (squares).}
\label{fig:rotsymblock}
\end{figure}

The constants $c_i$ may be chosen freely (except that
$\sum_j\omega(j)=1$ must be fulfilled). Now the blocking overlaps,
i.e.~a fine lattice point contributes to more than one point on the
new blocked lattice (figure \ref{fig:rotsymblock}). In momentum space,
the blocking function is given by
\begin{equation}
\omega(q)=c_o+\sum_{n=1}^d2^nc_n\sum_{\mu_1=1}^d\sum_{\mu_2=\mu_1+1}^d\cdots
\qquad\sum_{\mu_n=\mu_{n-1}+1}^d
\cos(q_{\mu_1})\cos(q_{\mu_2})\cdots\cos(q_{\mu_n})\label{overlapomega}
\end{equation}
We require the constants $c_i$ to be positive
since we do not want to have points with a negative weight:
 \[0\leq c_n\leq 2^{-n}\left(\begin{array}{l}d\\n
\end{array}\right)^{\!\!-1}\]
General simplifications concerning $\hat\omega$ and $\Omega$ only
exist for very specific $c_i$ values. If we choose the blocking to be
`flat', i.e.~the overall weight of the points to be equal, we get very
short-ranged action and a fast convergence which is most welcome in the
procedure of constructing a FP action numerically. For this flat
blocking the numbers $c_i$ are given by $c_i=2^{-d-i}$
where $i=0\ldots d$. Then, similarly to the Kadanoff blocking, we find
in momentum space by simplifying (\ref{overlapomega})
\begin{equation}
\Omega^{(k)}(Q)=\prod_\nu
\left(\frac{\sin(Q_\nu/2)}{2^k\sin(Q_\nu/2^{k+1})}\right)^{\!\!\! 4}
\label{rotOmega}
\end{equation}
which is identical to (\ref{wilsOmega}) up to the exponent.
Also the FP functions may be given explicitly.

\section{Optimizations}\label{optimizations}

\subsection{The Range of the Kadanoff-blocked Action in 1$d$}

For Kadanoff-blocked FP fermions in 1$d$ it is possible to choose the
constant $\kappa$ in such a way that the FP action becomes so
short-ranged that only nearest-neighbor interactions prevail. The
action is calculated by Fourier-transforming the inverse of the
propagator.
In one dimension, the summation over $\ell$ in eq.~(\ref{alphak},
\ref{betak}) can be written in a closed
form, yielding
\begin{eqnarray}
\alpha^{(1)}(Q)\!\! &\!\! =\!\! &\!\! 
\sin(Q)\cdot\frac{\sinh^2({m\over 2})}{m^2[\sin^2({Q\over
2})+\sinh^2({m\over 2})]}
\label{wil1da}\\
\beta^{(1)}(Q)\!\! &\!\! =\!\! &\!\!\frac{2}{\kappa}+\frac{m-\sinh(m)}{m^2}+\sinh(m)\cdot
\;\frac{\sinh^2({m\over 2})}{m^2[\sin^2({Q\over 2})+\sinh^2({m\over 2})]}
\label{wil1db}
\end{eqnarray}
The action consists only of
nearest-neighbor terms in $\rho^{(1)}(Q)$ and $\lambda^{(1)}(Q)$ if we
choose $\kappa$ to be
\begin{equation}
\kappa = \frac{2m^2}{e^m-m-1}\label{kapwils1d}
\end{equation}
For $m=0$ we have $\kappa=4$. Using this expression for $\kappa$ 
we have for the action the same form as the standard Wilson lattice
action for free fermions
but with different coefficients
\begin{eqnarray}
S^{(1)}[\bar{\Psi},\Psi]\!\! &\! =\! &\!\! \sum_{n,k\in
Z}\bar{\Psi}_\alpha(n)\,\Delta^{(1)}_{\alpha\beta}(n-k)\,\Psi_\beta(k)\label{act1d}\\
\nonumber \\
\Delta^{(1)}_{\alpha\beta}(n)\!\! &\! =\! &i\varrho^{(1)}(n){\mathbf{\gamma}_0}_{\alpha\beta}\,+\,\lambda^{(1)}(n)\cdot{\mathbf{\delta}_{\alpha\beta}}\\
\nonumber \\
i\varrho^{(1)}(n)\!\! &\! =\! &\frac{c(m)}{2}\Bigl(\delta(n+1)-\delta(n-1)\Bigr)
\label{rho1d}\\
\nonumber \\
\lambda^{(1)}(n)\!\! &\! =\! &d(m)\delta(n)+\frac{c(m)}{2}\Bigl(2\delta(n)-\delta(n+1)-\delta(n-1)\Bigr)\label{lambda1d}
\end{eqnarray}
where
\begin{eqnarray}
c(m)&=&\frac{\left(\frac{m}{2}\right)^2}{\sinh^2\left(\frac{m}{2}\right)}\,e^{-m}\\
\nonumber\\
d(m)&=&(e^m-1)c(m)
\end{eqnarray}
For $m=0$ we have $c(m)=1$ and $d(m)=0$, so we get back the Wilson
action with $r=1$.
Such a perfect optimization is only possible in one dimension. In
higher dimensions it seems that the value of $\kappa$ for the
optimally short-ranged action is still given by
(\ref{kapwils1d}). This has been shown also in \cite{wiesnew}.

\subsection{The Overlap-blocking in $1d$}\label{overlap1d}

In one dimension, the overlap blocking function given by
(\ref{overlapomega}) is just
\begin{equation}
\omega(q)=1-4c\,\sin^2(q/2)\label{omega1d}
\end{equation}
where $c$ is a constant $0\leq c\leq \frac{1}{2}$ we can play with.  We
start from some action which reproduces the correct continuum action in
the lattice spacing $\rightarrow 0$ limit. Most generally, such an
action may contain an infinite number of coupling constants, see
eq.~(\ref{genericS}).  We measure
the next-to-leading eigenvalue of the RG transformation, which is
in eq.~(\ref{nearfp}) is the second largest
irrelevant eigenvalue $e_\alpha$ (the largest eigenvalue is $1$
with the eigenoperator $S^\ast[\bar{\Psi},\Psi]$). If we have done many steps
and are near enough to the FP, all other smaller eigenvalues will have
so much smaller contributions that we can see only the remaining (larger)
eigenvalue and its operator. 

For the Kadanoff-blocking the eigenvalues are the same as one expects from
dimesional analysis-considerations. 
In the case of the overlap-blocking, the picture is surely
different since for $c=0,
\kappa=\infty$ we get back decimation, and there the eigenvalue was $1$
instead of $1/2$.

\begin{figure}[ht]
\begin{center}
\leavevmode
\epsfxsize=70mm
\epsfbox{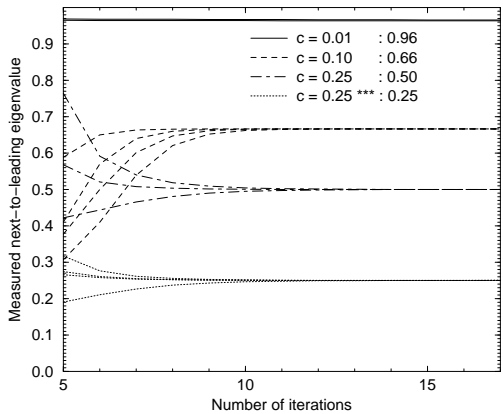}
\vskip 5mm
\end{center}
\caption{\em The determination of the next-to-leading eigenvalue from
$R^{(k)}(n)$, $L^{(k)}(n)$ at different values of $c$. The different curves
approaching the same limit correspond to different values of the distance
$n$. The eigenvalue
does not depend on $\kappa$ at all. However, it can happen that the
starting action lies on a line where the coefficient of the operator to
the next-to-leading eigenvalue is zero, and then we measure the
next-to-next-to-leading eigenvalue. The situation is illustrated
at $c=0.25$: for some general starting action the measured
next-to-leading-eigenvalue is $0.5$, but if we start from the
Wilson-fermion action at $r=1$ (***), we measure $0.25$ -- this is the
next-to-next-to-leading eigenvalue.}
\label{fig:lamfixconv}
\end{figure}

We try to measure the next-to-leading eigenvalue for each pair of
$(c,\kappa)$. In the discussion of the coupling constant space in
section \ref{eigencucc} we defined the linearized RGT in
(\ref{linRGT}). Here we consider many couplings simultaneously since the
two operators $\sum_i\gamma_\mu\partial_\mu\cdot a_i\hat{\Box}^i$ and $\sum_i
b_i\hat{\Box}^i$ ($i$ is running from 0 to infinity) corresponding to
$\rho(n)$ and $\lambda(n)$, respectively, include many possible
couplings $a_i, b_i$ ($\hat{\Box}$ is a symbolical form for the lattice
laplace operator). Couplings not included here may come in during the
blocking process. In order to measure the eigenvalue of interest, consider
the quantities
\begin{eqnarray}
R^{(k)}(n)&\, =\,& \frac{\rho^{(k+1)}(n)-\rho^{(k)}(n)}{
\rho^{(k)}(n)-\rho^{(k-1)}(n)}\label{Rk}\\
\nonumber\\
L^{(k)}(n)&\, =\,& \frac{\lambda^{(k+1)}(n)-\lambda^{(k)}(n)}{
\lambda^{(k)}(n)-\lambda^{(k-1)}(n)}\label{Lk}
\end{eqnarray}
where $k$ gives the number of blocking steps and $n$ stands for the
lattice site. If we are near enough to the FP we may write $\rho(n) =
\rho^\ast(n)+\Delta\rho(n)$ and $\lambda(n) = \lambda^\ast(n) + 
\Delta\lambda(n)$. Then in $R(n)$ and $L(n)$ the FP contributions
cancel.  If the next-to-leading eigenvalue is non-degenerate then
after sufficiently many blockings $k$, $R^{(k)}(n)$ and $L^{(k)}(n)$
will reach this eigenvalue. The procedure of finding the
second-largest eigenvalue is displayed in figure \ref{fig:lamfixconv}.

\begin{figure}[ht]
\begin{center}
\leavevmode
\epsfxsize=73mm
\epsfbox{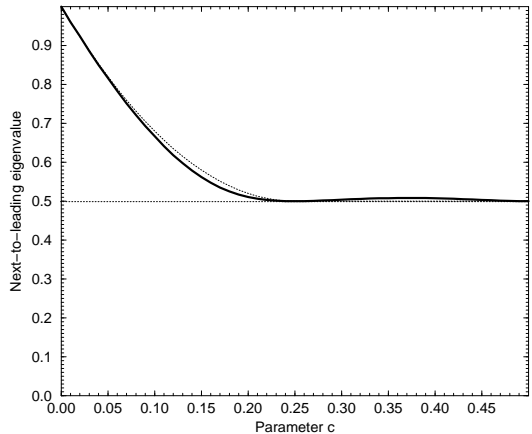}
\vskip 5mm
\end{center}
\caption{\em The next-to-leading eigenvalue of convergence 
depending on $c$. The dotted line is $c=1-4c+8c^2$; the prediction for small $c$.}
\label{fig:cconva4}
\end{figure}

What we find is displayed in figure
\ref{fig:cconva4}: the eigenvalue seems to be independent of $\kappa$,
and it is a smooth function of $c$.
The exact eigenvalue $1/2$ is only dominant for $c=0.25$ and $c=0.5$. For
every other $c$ it is larger; it is a redundant operator. 
For decimation the
redundant operator is obviously the one which is changing the parameter
$r$. We can write down how this operator changes the FP of decimation
(\ref{decrho}, \ref{declambda}) by adding a small perturbation
to the Wilson parameter $r\rightarrow r+\varepsilon$.
Since such a perturbed action does not change the behaviour of the action 
under the RG transformation for
decimation (it also represents a FP), the difference $\delta\rho$ and
$\delta\lambda$ remains the same for the next step.  The next-to-leading
eigenvalue is therefore exactly $1$.

Assuming a perturbation along an eigenvector $\Delta_1(q)$ of the RG
transformation with eigenvalue $e$,
\begin{equation}
\Delta(q)=\Delta^\ast(q)+\varepsilon\Delta_1(q)
\end{equation}
with $\varepsilon\ll 1$, one obtains from eq.~(\ref{Sonestep}) in the linear
approximation 
\begin{equation}
e\cdot{\Delta^\ast}^{-1}(Q)\,
\Delta_1(Q)\,{\Delta^\ast}^{-1}(Q)\,=\,\frac{1}{2}
\sum_{\ell=0}^1\,{\Delta^\ast}^{-1}\left(
\frac{Q}{2} +\pi\ell\right)\,
\Delta_1\left(
\frac{Q}{2} +\pi\ell\right)\label{deltadelta}\label{dede}
\end{equation}
\[
\qquad\qquad\,{\Delta^\ast}^{-1}\left(
\frac{Q}{2} +\pi\ell\right)\cdot\left|\,\omega\left(\frac{Q}{2}
+\pi\ell\right)\right|^2
\]
Since 
\begin{equation}
\Delta^\ast(q)\;\; =\; \gamma_\mu\rho^\ast_\mu(q)+\lambda(q)\;\;=\;\gamma_\mu q_\mu
+O(q^2),
\end{equation}
the eigenoperator with the leading eigenvalue should have the $q\rightarrow 0$
behaviour given by
\begin{equation}
\Delta_1(q) = q^2+O(q^3)\label{redop}
\end{equation}
Evaluating eq.~(\ref{dede}) in the $Q\rightarrow 0$ limit one has
\begin{equation}
e=\frac{1}{2}+\frac{1}{2}\sum_{\ell\neq
  0}\Delta^\ast(\pi\ell)\Delta_1(\pi\ell){\Delta^\ast}^{-1}(\pi\ell)\left|\omega(\pi\ell)\right|^2
\end{equation}
This expression shows that in general, for $\omega(\pi\ell)\neq 0$, the
corresponding eigenvalue $e\neq 1/2$. For small values of $c$ one can estimate
the eigenvalue as follows. At $c=0$ (decimation) $\omega(q)=1$ and in this
casse $\Delta_1(q)=(\Delta^\ast(q))^2$ is clearly a solution to
eq.~(\ref{dede}) with $e=1$. Taking this form of $\Delta_1(q)$ approximately
valid for $c\neq 0$ one obtains
\begin{equation}
c=\frac{1}{2}+\frac{1}{2}\sum_{\ell\neq  0}\left|\omega(\pi\ell)\right|^2
\end{equation}
In $d=1$, this gives $e=1-4c+8c^2$. Figure \ref{fig:cconva4} shows taht this
estimate agrees well with the observed leading eigenvalue. Note, however, that
a perturbation to the FP actoin which is $\propto q^2$ does not result in an
artifact in the spectrum, since it vanishes on the mass shell
$q^2=0$. Therefore this eigenvalue $e\geq 1/2$ corresponds to a redundant
direction.

Luckily, the operators responsible for the lattice artifacts vanish much
faster than the redundant operator discussed above. If we choose
$\Delta_1(q)= \sum_\mu\gamma_\mu q^3_\mu + O(q^4)$
 then in equation (\ref{dede}) the
l.h.s.~has a singularity at $q=0$ (unlike in the case of (\ref{redop}) 
). On the r.h.s.~only the $\ell=0$ term can
give such singular terms, and it is easy to show that the corresponding
eigenvalue is $1/4$. For $\Delta_1(q)=\sum_\mu q_\mu^4 + O(q^5)$ the
eigenvalue can be calculated similarly, the result is $e=1/8$. These two
directions correspond to the two leading operators 
\[a^2\int d^dx \sum_\mu\bar{\Psi}\gamma_\mu\partial_\mu^3\Psi\quad;\quad a^3\int
d^dx\sum_\mu\bar{\Psi} \partial_\mu^4\Psi\] 
in the expansion of the lattice action in powers of the
lattice spacing $a$. These are responsible for the largest artifact
contributions.  For such operators the dimensional analysis yields again the
correct eigenvalues. So operators giving rise to lattice artifacts do not
have large eigenvalues in the RG transformations.

\subsection{The interaction range in $1d$}

In the context of improved lattice actions, we want to have a FP action
which is as short ranged as possible. Since the Kadanoff blocking leads to a perfectly local action, we hope
so will this blocking for appropriate $\kappa$ and $c$- values. Of
course we already found these values: the decimation $c=0$,
$\kappa=\infty$ gives a nearest-neighbor action for $r=1$ -- the same
as the Kadanoff-blocked action at $\kappa=4$. But as we have mentioned, 
decimation is not a
very good RG transformation to consider.

\begin{figure}[ht]
\begin{center}
\leavevmode
\epsfxsize=70mm
\epsfbox{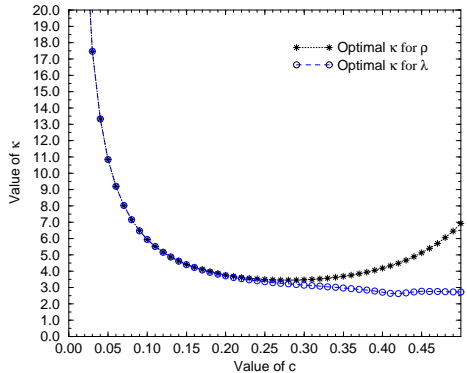}
\end{center}
\caption{\em The optimal $\kappa$ values depending on $c$. For $c>0.2$,
the optimal $\kappa$ is not the same for $\lambda(n)$ and $\rho(n)$.}
\label{fig:optimkc}
\end{figure}
 Moreover, its largest and
second-largest eigenvalues are both $1$, and slow convergence is 
unwanted in computer simulations.  So we try to find other
short-ranged actions for different pairs of $(c,\kappa)$.  The
short-range property of the action seems to depend largely on
$\kappa$, so we optimize $\kappa$ for every value of $c$.  The result
is plotted in figure \ref{fig:optimkc}.

\begin{figure}[ht]
\begin{center}
\leavevmode
\epsfxsize=70mm
\epsfbox{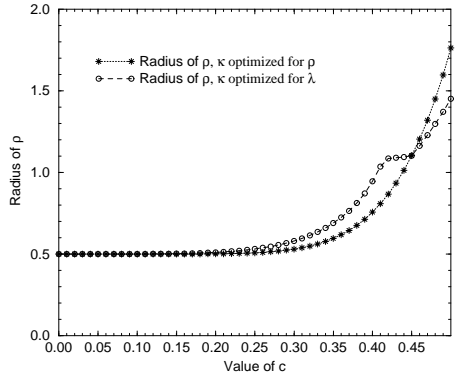}
\end{center}
\caption{\em The radius for the optimal $\kappa$ values depending on
$c$.}
\label{fig:radiusrho}
\end{figure}

Up to the first half of the $c$-values, the optimized action
remains fairly local. It is a smooth transition between the
nearest-neighbor action $c=0$, $\kappa=\infty$, and the not very short
ranged actions at $c>0.3$. Figure \ref{fig:radiusrho} shows the radius
of $\rho(n)$ defined by 
\begin{equation}
r_\rho^2 = \frac{\sum_n|\rho(n)|n^2}{\sum_n|\rho(n)|}\label{rhorad}
\end{equation}

For $c<0.3$ and the corresponding 
optimal $\kappa$ the value of $r_\rho$ is quite small. At small values of $c$,
the next-to-leading eigenvalue is close to $1$, therefore it seems that the
best choice is found in the middle of the possible $c$-values: For
$c=0.25$ the action is still very local and it converges very
fast. Additionally, it may be treated analytically. Actions with
$c<0.25$ are a little bit more short-ranged, but their eigenvalue of
convergence is larger and there are no nice analytic expressions for
them.

\subsection{Optimal action for 4$d$ including the mass}\label{optimkappa4d}

We assume that in $4d$ the situation is similar to the one
dimensional case. The 'flat' blocking seems to be one of the
optimally short ranged actions, moreover it is treatable analytically
much further than for other choices of $c_i$. The optimal $\kappa$ is
$\kappa=3.3675$ for the massless action if we optimize for $r_\rho$.
The value of $r_\rho$ stays at roughly $r_\rho=1.6$ for all masses where
optimization is possible.

We may now extend the optimization to the massive actions.
The parameter $\kappa$ may be optimized at each step on the renormalized
trajectory. We first move into the Gaussian FP and then add a small
mass, say $m=0.001$. Now at each blocking the value of $\kappa$ is
optimized for locality. The result of the optimization is displayed
below in figure \ref{fig:optimmass}.
\begin{figure}[ht]
\begin{center}
\leavevmode
\epsfxsize=70mm
\epsfbox{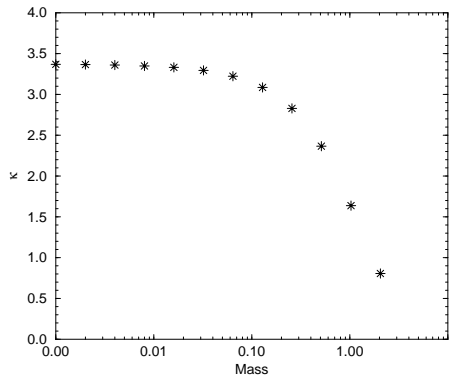}
\end{center}
\caption{\em The optimal values of $\kappa$ for mass values up to
$m=2$. We see that the optimal $\kappa$ values would approach zero for
masses larger than $m=3$. For $m>2$ an optimization by tuning $\kappa$
does not lead to a sufficiently short ranged action.}
\label{fig:optimmass}
\end{figure}

\section{Conclusions}
The convergence properties of a general RG transformation have been
analized in this work. The observation is that the next-to-leading
eigenvalue of the transformation can be in principle any number
depending on the choice of the blocking function. 
This eigenvalue corresponds to a redundant operator, which does not produce
lattice artifacts. The eigenvalues obtained by dimensional analysis are still
present -- they are responsible for the artifacts $a^{2n}$.
We have shown that it is possible to optimize the parameters of the
block transformation for locality and that 'flat' blocking
configurations where the fine lattice sites are weighted equally to
obtain the coarse fields are preferred candidates for local actions.

A more detailed discussion can be found in \cite{diss}.
I want to thank Ferenc Niedermayer for many useful discussions and Peter
Hasenfratz for his support.

\end{document}